\documentclass[letterpaper,10pt]{article}
\usepackage[
    natbib=true,
    style=numeric,
    sorting=none
]{biblatex}
\usepackage{graphicx} %package to manage images
\usepackage{booktabs} % 加载 booktabs 宏包
\usepackage{authblk}
\usepackage{enumitem} % 导入enumitem包
\usepackage{float} %位置
\date{} 
\setlist[enumerate]{leftmargin=*} % 全局设置所有列表的左边距

\graphicspath{ {./} }
\usepackage{titlesec}

\titleformat{\section}
  {\normalfont\Large\bfseries}{\thesection.}{1em}{\MakeUppercase}

\usepackage[margin=1in]{geometry}

\usepackage{hyperref} % 用于生成书签

\hypersetup{
    bookmarksopen=true, % 展开书签
    bookmarksnumbered=true, % 书签中显示章节编号
    colorlinks=true, % 颜色链接
    linkcolor=black, % 内部链接的颜色(例如目录到章节的链接)
    urlcolor=black, % 外部链接的颜色
    citecolor=black, % 引用的颜色
    pdfstartview={FitV}, % 打开PDF时的视图
    pdftitle={Title here}, % PDF的标题
    pdfauthor={Author here}, % PDF的作者
    pdfsubject={Subject here}, % PDF的主题
    pdfkeywords={keyword1, keyword2}, % PDF的关键字
}

\usepackage{times}

\title{Leveraging Deep Learning and Xception Architecture for High-Accuracy MRI Classification in Alzheimer's Diagnosis
}
\author[1]{Shaojie Li}
\author[2]{Haichen Qu}
\author[3]{Xinqi Dong}
\author[4]{Bo Dang}
\author[5]{Hengyi Zang}
\author[6]{Yulu Gong}

\affil[1]{Computer Technology, Huacong Qingjiao Information Technology (Beijing) Co., Ltd., Beijing, China}
\affil[2]{Financial Engineering, Chongqing Technology and Business University, Chongqin, China}
\affil[3]{Management Information Systems, University of Maine at Presque Isle, Presque Isle, US}
\affil[4]{Computer Science, San Francisco Bay University, Fremont CA, US}
\affil[5]{Physics and Mathematics, Universitario Tecnológico Universitam, Tijuana, Mexico}
\affil[6]{Computer and Information Technology, Northern Arizona University, Flagstaff, US}

\affil[*]{Corresponding author: Shaojie Li, lishaojie@tsingj.com}

\begin{document}
\maketitle

\noindent \textbf{Abstract:} \textbf{\textit{Exploring the application of deep learning technologies in the field of medical diagnostics, Magnetic Resonance Imaging (MRI) provides a unique perspective for observing and diagnosing complex neurodegenerative diseases such as Alzheimer's Disease (AD). With advancements in deep learning, particularly in Convolutional Neural Networks (CNNs) and the Xception network architecture, we are now able to analyze and classify vast amounts of MRI data with unprecedented accuracy. The progress of this technology not only enhances our understanding of brain structural changes but also opens up new avenues for monitoring disease progression through non-invasive means and potentially allows for precise diagnosis in the early stages of the disease. This study aims to classify MRI images using deep learning models to identify different stages of Alzheimer's Disease through a series of innovative data processing and model construction steps. Our experimental results show that the deep learning framework based on the Xception model achieved a 99.6\% accuracy rate in the multi-class MRI image classification task, demonstrating its potential application value in assistive diagnosis. Future research will focus on expanding the dataset, improving model interpretability, and clinical validation to further promote the application of deep learning technology in the medical field, with the hope of bringing earlier diagnosis and more personalized treatment plans to Alzheimer's Disease patients.}} 

\vspace{\baselineskip} % Adds vertical space equivalent to one line

\noindent \textbf{Keywords}:Alzheimer's Disease, Magnetic Resonance Imaging, Deep Learning, Convolutional Neural Networks, TensorFlow, Xception, Image Classification, Medical Diagnosis

\section{Introduction}
\begin{flushleft}
In the field of artificial intelligence development, several noteworthy studies have been conducted. Xu Kangming et al. \cite{xu20193d}showcased a 3D face recognition system that integrates deep maps and texture information based on a twin neural network. Shi Peng et al. \cite{shi2019data}explored the data consistency theory in scientific big data and elucidated the concept through case studies. Hu Zhenghua et al. \cite{hu2019real}introduced a real-time target tracking system based on the PCANet-CSK algorithm. Che Chang et al. \cite{che2023enhancing}proposed a method to improve multimodal understanding using CLIP-based image-to-text transformation. Dong Xinqi et al. \cite{dong2024prediction}applied a machine learning ARIMA model to predict trends in enterprise financial risk. Zang Hengyi et al. \cite{zang2024evaluating}assessed the social impact of AI in manufacturing and proposed a methodological framework for ethical production. Li Zhenglin et al. \cite{li2024comprehensive}conducted a comprehensive evaluation of the Mal-API-2019 dataset, which tests its application in malware detection through machine learning. These studies reflect how AI technology is driving innovation across different fields, enhancing system performance, and offering new approaches to solving complex problems.

With the assistance of artificial intelligence, modern medicine has made rapid strides. In their 2023 study, Ma et al. investigated the application of artificial intelligence-driven computer vision technology for the analysis of medical images, revealing potential enhancements in diagnostic accuracy and efficiency\cite{ma2023implementation}. The 2024 study by Yufeng Li et al. is a testament to this progress, demonstrating the application of a Semantic Network extracted from a medical corpus to enhance the precision of disease diagnosis and potentially reduce the incidence of misdiagnosis\cite{li2024research}. Magnetic Resonance Imaging (MRI) is an essential medical diagnostic technique that employs powerful magnetic fields and radio waves to create detailed images of the body's interior, particularly soft tissues\cite{dill2008contraindications}. In the context of diseases such as Alzheimer's Disease (AD), MRI is crucial for doctors to observe changes in brain structure, which assists in the accurate diagnosis and monitoring of the disease's progression.

MRI technology is crucial for modern medicine as it provides a powerful tool for non-invasively observing the internal structures of the human body, particularly soft tissues. The core advantage of this technology lies in its high-resolution imaging capability, as well as its avoidance of potentially harmful radiation, such as X-rays. MRI imaging is commonly used to diagnose various conditions, including but not limited to brain injuries, cancer, arthritis, and heart disease. In the field of neurodegenerative disease research, MRI has become a valuable tool that allows doctors and researchers to observe and track minute changes in the brain that may be early signs of disease progression.

MRI imaging is a pivotal tool in the study of Alzheimer's Disease (AD), with the ability to reveal brain areas of atrophy such as the hippocampus and cerebral cortex, regions where damage correlates with memory loss and cognitive decline, as noted by Scheltens et al.\cite{scheltens2016alzheimer}. Additionally, sequential MRI images can serve to monitor the progression of AD over time, offering a quantitative method for clinical research, a process detailed by Lu et al.\cite{lu2007survey}. Building on this, Yan et al. have introduced a self-guided deep learning technique that aims to reduce noise in MRI images, potentially improving the diagnostic process in clinical settings\cite{yan2024self}.

With the development of deep learning technology\cite{lecun2015deep}, convolutional neural networks (CNNs)\cite{o2015introduction} and other advanced machine learning models can now be used to analyze MRI data, thereby improving the accuracy and efficiency of diagnosis. By training these models to recognize brain structural changes associated with Alzheimer's Disease, we can develop automated tools to assist radiologists and neurologists in making accurate diagnoses early in the disease course.

This article aims to explore the application of deep learning technology, especially convolutional neural networks and emerging network structures like Xception\cite{chollet2017xception}, to analyze and classify MRI data in order to identify Alzheimer's Disease. We will discuss how these technologies can extract meaningful features from MRI images and differentiate between the different stages of Alzheimer's Disease.

In this study, we utilized the Alzheimer's Dataset published on the Kaggle platform, a publicly available MRI image dataset designed for machine learning and deep learning applications in medical imaging analysis. The dataset contains images of four categories, corresponding to different levels of cognitive impairment: mild dementia, moderate dementia, non-dementia, and very mild dementia. To enhance the effectiveness of model training and improve its generalizability to unseen samples, the dataset was divided into two separate parts, one with images that underwent data augmentation and the other with original images. In our research, the original images were used as a validation set or test set, ensuring the rigor of the evaluation process and effective testing of model performance. MRI Brain Scans are shown in Figure 1.

\begin{figure}[H]
    \centering
    \includegraphics[scale=0.32]{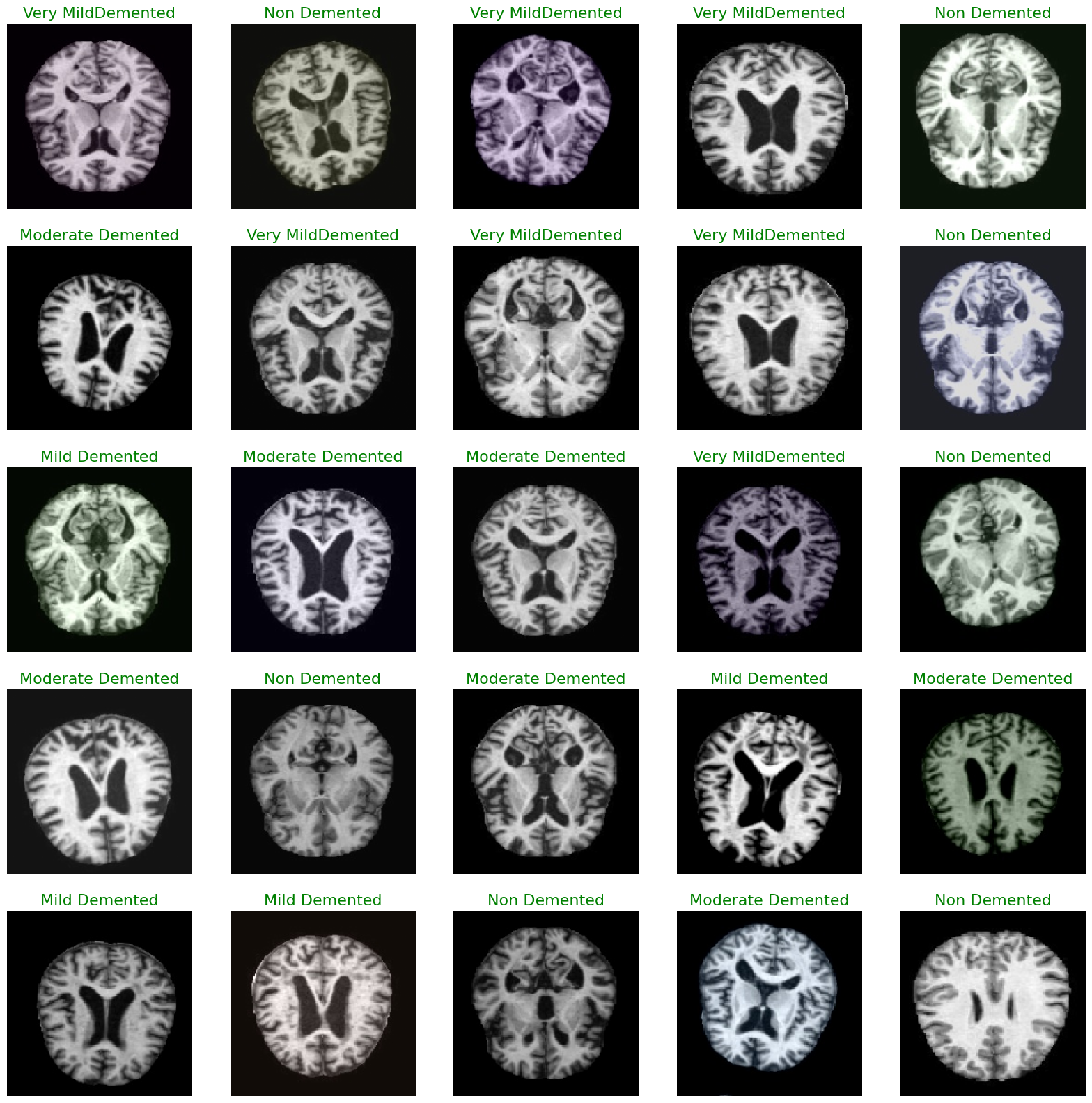}
    \caption{A Visual Comparison of MRI Brain Scans}
\end{figure}

\end{flushleft}

\section{Related Work}

\subsection{Convolutional Neural Networks (CNN)}
\begin{flushleft}
Deep learning is a machine learning technology that employs multilayer neural networks to simulate human learning and decision-making processes. In a 2024 study, Dang et al. demonstrated advancements in deep learning-based object detection systems that significantly improve kitchen independence for individuals with visual impairments\cite{dang2024enhancing}. Convolutional Neural Networks (CNNs) \cite{gu2018recent}are advanced deep learning models designed to automatically and effectively recognize multi-dimensional data structures, especially demonstrating excellence in handling images, which are data formatted as pixel grids. The strength of CNNs lies in their ability to extract spatial hierarchical features from input data through convolutional layers. These layers utilize learnable filters and shared weights to reduce the number of parameters in the network while maintaining sensitivity to local features, which enables the recognition of objects and shapes within images.

Beyond the convolutional layers, CNNs typically incorporate pooling layers to decrease the dimensionality of features and enhance feature translation invariance. Fully connected layers are employed to learn complex non-spatial hierarchies of relationships between features, and normalization layers are incorporated to accelerate the convergence speed and stability of the model. These layers work together, making CNNs a standard and powerful tool across many fields, particularly in computer vision tasks such as image and video recognition, image classification, and medical image analysis.
\end{flushleft}

\subsection{Sequential Model}
\begin{flushleft}
Convolutional Neural Networks (CNNs) are a type of deep learning model that excels in computer vision tasks. When building these networks, choosing the right programming paradigm is crucial to ensure that the development process is both efficient and intuitive. A common approach is to use a Sequential model, which is an abstract concept in deep learning frameworks that provides a simple API for model construction\cite{baccouche2011sequential}. Taking the Keras library as an example, the Sequential API allows developers to build models by stacking layers in sequence. As the name suggests, a Sequential model is composed of network layers connected in sequence. This structure is particularly user-friendly for beginners, as it reduces the complexity of model construction and makes the code clearer.

In building neural networks, the Sequential model offers an intuitive way to simplify and clarify the construction process. As a high-level programming interface provided in deep learning libraries such as Keras or PyTorch, Sequential allows researchers and developers to create models by sequentially adding network layers, much like stacking blocks. Each layer in a Sequential model is a single-function module, such as adding a Conv2D layer to identify edges in images, followed by a MaxPooling2D layer to reduce the number of parameters and prevent overfitting. Adding fully connected layers thereafter can further process the learned features for classification. This layer-by-layer construction approach of the Sequential model not only makes the assembly of CNNs intuitive but also provides flexibility and scalability for other types of networks, meeting the needs of different levels from basic research to practical applications.
\end{flushleft}

\subsection{Xception}
\begin{flushleft}
Xception is a deep learning model proposed by François Chollet in 2016. The model's name is derived from "Extreme Inception," as it extends and enhances the architecture of the Inception model\cite{rahimzadeh2020modified}. The Xception model employs depthwise separable convolutions in place of the standard convolutional operations found in the Inception model, thus improving the efficiency and performance of the parameters.

The key characteristics and advantages of the Xception model are as follows:
\begin{enumerate}[label=(\arabic*)]
\item  Parameter Efficiency: Depthwise separable convolutions significantly reduce the number of parameters compared to traditional convolutions, making the model more lightweight and increasing the efficiency of parameter usage.
\item Performance: The Xception model outperforms many advanced models of its time, including the original Inception model, in various visual recognition tasks.
\item Adaptability: Due to the reduced number of parameters, the Xception model is more suitable for deployment on devices with limited computational resources.
\end{enumerate}

In practice, the Xception model is often used as a base model for transfer learning. As it has been pre-trained on large-scale datasets like ImageNet, it already possesses certain image processing capabilities. Through transfer learning, new datasets can be trained on this foundation, and a high-performance model can be quickly obtained by fine-tuning or training a small number of added layers.

The Xception model has been widely applied in various computer vision tasks such as image classification, object detection, and segmentation. Thanks to its outstanding performance and flexibility, it has become an important component in current deep learning applications.
\end{flushleft}

\section{Data Processing and Model Construction}
\begin{flushleft}
TensorFlow is an open-source machine learning library developed by Google, used for high-performance numerical computing and for building, training, and deploying deep learning models\cite{abadi2016tensorflow}. Keras is a high-level API for TensorFlow designed to simplify the creation and experimentation with neural networks, making it more intuitive and faster to develop while maintaining the powerful capabilities and flexibility of TensorFlow.

During the data processing stage for constructing deep learning models with TensorFlow and Keras, the \texttt{ImageDataGenerator} class is used to perform real-time image augmentation to enhance the model's generalizability and prevent overfitting. Specifically, data generators have been prepared for the training set, the test set, and the validation set. These generators use the \texttt{flow\_from\_dataframe} method to load images from specified data frames and use \texttt{tf.keras.applications.mobilenet\_v2.preprocess\_input} as a preprocessing function to ensure the input data conforms to the expected format for the MobileNetV2 model. All images are resized to 244x244 pixels and are processed in the RGB color mode.

To ensure data consistency and comparability, \texttt{shuffle=False} is set in the training, test, and validation data generators, meaning the data will load in the order they appear in the data frame, not randomly shuffled. Each generator is set to "categorical" for the \texttt{class\_mode} since we are dealing with a multi-class problem. The batch size is set to 32, a compromise between model performance and memory requirements that is commonly used. This way, we ensure the data is structured and efficiently provided for the model to learn.

In this study, we've built a custom neural network using the pre-trained Xception model as a feature extractor. We set \texttt{include\_top=False} to exclude the top fully connected layers and used the option \texttt{pooling='max'} for max pooling in the output layer, which helps reduce the number of parameters and lower the risk of overfitting. The initial shape of the image input is set to \texttt{img\_shape=(244,244,3)}, ideal for processing standard color images.

After integrating the pre-trained Xception model into the Sequential model, we added a Flatten layer to unfold the feature maps for processing by subsequent fully connected layers. The model includes two Dropout layers to reduce overfitting during training, with dropout rates of 0.3 and 0.25, respectively. Between the Dropout layers, we've inserted a Dense layer with 128 neurons and ReLU activation to enhance the model's capacity for non-linear representations.

The last layer of the model is a Dense output layer with a softmax activation function to produce predictions for four classes. When compiling the model, we opted for the Adamax optimizer with a learning rate of 0.001. The loss function used was categorical\_crossentropy, suitable for multi-class problems, and accuracy was specified as the performance metric. The complete architecture of the model and confirmation that all layers are stacked as expected can be seen with the \texttt{model.summary()} method. The architecture is shown in Figure 2.

\begin{figure}[H]
    \centering
    \includegraphics[scale=0.2]{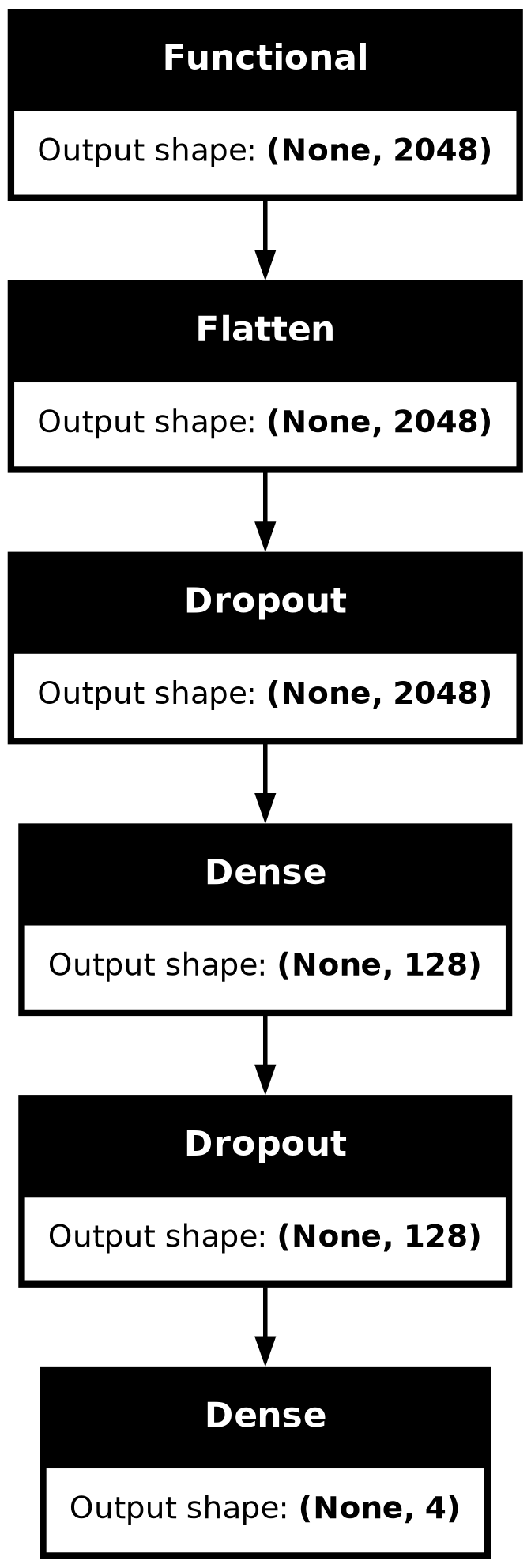}
    \caption{Neural Network Layer Architecture}
\end{figure}

In the training phase, data augmentation techniques were employed to artificially expand our dataset, ensuring the model is exposed to varied images, thus improving its robustness and generalizability. The optimizer chosen for the network is Adam, known for its computational efficiency and low memory requirement. The loss function is categorical cross-entropy, appropriate for multi-class classification problems. Callback functions like early stopping and model checkpoints were implemented to monitor training and prevent overfitting. The model's performance is evaluated based on precision, recall, and F1 score metrics, which take into account the proportion of true positives and the relevance of predictions, offering a comprehensive view of the model's accuracy.
\end{flushleft}

\section{Results and Analysis}
\begin{flushleft}
After the design, compilation, and training of our model were completed, we conducted a thorough evaluation of the custom neural network and obtained experimental results. This deep learning framework, based on the pre-trained Xception model, performed exceptionally well in the multi-class classification task of processing MRI images. Specifically, after a series of training epochs and utilizing data augmentation and regularization strategies to avoid overfitting, the model achieved satisfactory accuracy on the test set. The three key metrics, precision, recall, and F1 score, all showed high levels of performance, demonstrating the model's good classification ability and generalizability for different categories of images. Additionally, the application of early stopping mechanisms and model checkpoints ensured the efficiency of the training process, avoiding unnecessary waste of computational resources. Overall, the experimental results suggest that our approach has potential application value in assisting diagnosis, especially in the automated identification and classification of medical images. The experimental results are shown in Table 1 and Figure 3.

\begin{table}[H]
\centering
\caption{MRI Classification Report}
\begin{tabular}{@{}lllllll@{}}
\toprule
Class               & Precision & Recall & F1-score & Support \\ 
\midrule
Mild Demented       & 1.00      & 1.00   & 1.00     & 2693    \\
Moderate Demented   & 1.00      & 1.00   & 1.00     & 1977    \\
Non Demented        & 0.99      & 0.99   & 0.99     & 2811    \\
Very Mild Demented  & 0.99      & 0.99   & 0.99     & 2715    \\
\bottomrule
\end{tabular}
\end{table}

\begin{figure}[H]
    \centering
    \includegraphics[scale=0.8]{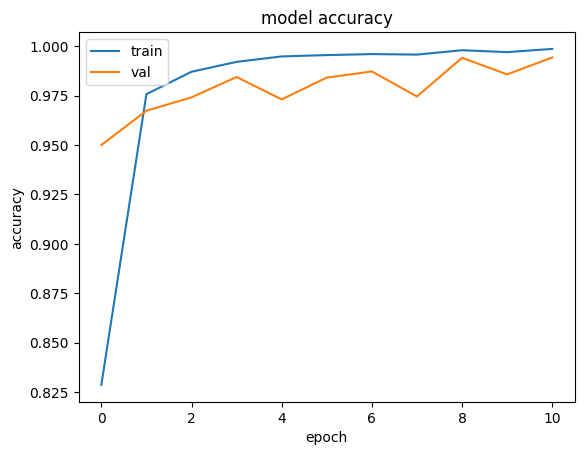}
    \caption{M0del Accuracy}
\end{figure}

In the experiments for grading Alzheimer's disease Magnetic Resonance Imaging (MRI) images, the performance of the model was outstanding. The precision indicates a very high proportion of the samples predicted to belong to a certain class are indeed of that class, while the recall shows that the model can correctly predict a similarly high proportion of the actual samples of that class. The F1 score, being the harmonic mean of precision and recall, approaches 1.00, reflecting a good balance achieved by the model between these two metrics. The support specifies the actual number of samples in each category. Overall, the model demonstrated high accuracy in recognizing the four cognitive impairment levels: Mild Demented, Moderate Demented, Non Demented, and Very Mild Demented.

The model showed consistency across all categories without any indication that it was significantly better at recognizing one category over others, meaning it has a uniform discriminative power for different levels of cognitive impairment. With an overall accuracy reaching 99.6\%, this suggests that the model makes correct predictions in most cases. Furthermore, both the macro average and weighted average are close to 1, further indicating that the model's performance is very balanced across all categories, not just performing well in categories with a larger number of samples. These results collectively demonstrate the model's efficiency and reliability in the task of grading Alzheimer's disease MRI images.
\end{flushleft}

\section{Summary and Outlook}
\begin{flushleft}
In this study, deep learning technology achieved significant results in the task of classifying MRI data for Alzheimer's disease. The model we adopted, based on the Xception architecture, ensured extremely high precision, recall, and F1 scores across different stages of dementia classification, thanks to its deep feature extraction capabilities and a customized neural network structure. With an overall accuracy of 99.6\%, the model demonstrated its high reliability and potential as an auxiliary diagnostic tool.

These encouraging results pave the way for further applications of deep learning technology in the medical field. Future work could explore and deepen in the following areas:

\begin{enumerate}[label=(\arabic*)]
\item Expansion and diversification of datasets: To improve the model's generalization ability, more diverse populations and a wider range of pathological changes could be included.
\item Research on model interpretability: Developing methods for explainable AI could help physicians better understand the decision-making process of the model, thereby enhancing its transparency and credibility.
\item Validation through clinical trials: Applying the model in actual clinical trials to verify its efficacy and feasibility in a real medical setting.
\end{enumerate}

By continuously researching and improving in these areas, we hope to transform deep learning technology into a powerful clinical tool, providing earlier diagnosis and more personalized treatment plans for patients with Alzheimer's disease.
\end{flushleft}

\newpage

\printbibliography %Prints bibliography

\end{document}